\documentclass{article}
\usepackage[utf8]{inputenc}

\usepackage{geometry}
\usepackage{float}
\usepackage{xcolor}
\usepackage{enumitem}
\usepackage{endnotes}
\usepackage{ulem}
\let\footnote=\endnote
\usepackage{titlesec}
\titleformat{\section}
{\normalfont\normalsize\bfseries}{\thesection}{1em}{}
\titleformat{\subsection}
{\normalfont\normalsize\itshape}{\thesubsection}{1em}{}

\usepackage[bitstream-charter]{mathdesign}

\usepackage{amssymb}
\usepackage{amsmath} 
\usepackage{mathtools}
\usepackage{array}
\usepackage{bbm} 
\usepackage{breqn} 
\usepackage{siunitx}

\newcommand{\norm}[1]{\left\lVert#1\right\rVert}

\usepackage{booktabs}
\usepackage{threeparttable}

\newcommand{\mc}[3]{\multicolumn{#1}{#2}{#3}}

\usepackage{subcaption}

\usepackage{appendix}
\usepackage{natbib}
\usepackage[hidelinks]{hyperref}
\usepackage{cleveref}

\begin{document}
\captionsetup{margin=10pt,font=footnotesize,labelfont=bf,labelsep=endash,justification=centerlast}

\begin{center}
    \LARGE
    \textbf{Exploration of the Parameter Space in Macroeconomic Models}\\[2ex]
\end{center}

\begin{center}
\small
\noindent \textbf{Karl Naumann-Woleske} (0000-0001-8182-9256)\\
Chair of EconophysiX and Complex Systems, Ecole Polytechnique Paris\\
LadHyX, UMR CNRS 7646, Ecole Polytechnique Paris\\
\href{mailto:karl.naumann@ladhyx.polytechnique.fr}{karl.naumann@ladhyx.polytechnique.fr}\\[2ex]
\textbf{Max Sina Knicker} (0000-0003-0732-6612)\\
Chair of EconophysiX and Complex Systems, Ecole Polytechnique Paris\\
Physics of Complex Biosystems, Technical University of Munich\\[2ex]
\textbf{Michael Benzaquen} (0000-0002-9751-7625)\\
Chair of EconophysiX and Complex Systems, Ecole Polytechnique Paris\\
LadHyX, UMR CNRS 7646, Ecole Polytechnique Paris\\
Capital Fund Management, 75007 Paris, France\\[2ex]
\textbf{Jean-Philippe Bouchaud} (0000-0002-0427-6456)\\
Chair of EconophysiX and Complex Systems, Ecole Polytechnique Paris\\
Capital Fund Management, 75007 Paris, France\\
Acad\'emie des Sciences, Quai de Conti, 75006 Paris, France\\[10ex]
\end{center}

\begin{center}
    \textbf{Abstract}
\end{center}

Agent-Based Models (ABM) are computational scenario-generators, which can be used to predict the possible future outcomes of the complex system they represent. To better understand the robustness of these predictions, it is necessary to understand the full scope of the possible phenomena the model can generate. Most often, due to high-dimensional parameter spaces, this is a computationally expensive task. Inspired by ideas coming from systems biology, we show that for multiple macroeconomic models, including an agent-based model and several Dynamic Stochastic General Equilibrium (DSGE) models, there are only a few \textit{stiff} parameter combinations that have strong effects, while the other \textit{sloppy} directions are irrelevant. 
This suggests an algorithm that efficiently explores the space of parameters by primarily moving along the stiff directions. We apply our algorithm to a medium-sized agent-based model, and show that it recovers all possible dynamics of the unemployment rate. The application of this method to Agent-based Models may lead to a more thorough and robust understanding of their features, and provide enhanced parameter sensitivity analyses. Several promising paths for future research are discussed. 
\newpage

\section{Introduction}\label{sec:introduction}

The global economy can be described as a complex adaptive system where large numbers of heterogeneous agents interact in parallel \citep{Tesfatsion2002, LeBaronTesfatsion2008a}.
The evolution of this system gives rise to emergent phenomena such as business cycles, fat tails in company sizes/agent wealths, or even crises and crashes. 
Agent-based models (ABMs) provide tools to generate and understand these emergent phenomena (for deeper introductions to agent-based modeling see \citet{Epstein1999}, \citet{LeBaronTesfatsion2008a}, \citet{DosiRoventini2019}, \citet{HaldaneTurrell2019}, \citet{KirmanGallegati2022}, and \citet{Arthur2022}).
Indeed, an agent-based model is a computational device for simulating artificial economies of many heterogeneous interacting agents. The richness of phenomena that an ABM can generate makes them excellent \textit{scenario-generators} \citep{GualdiEtAl2015, Bookstaber2017, Bouchaud2020} that can be used to understand the spectrum of possible paths for the economy, and how these future scenarios may respond to policy interventions. 
Unfortunately, the versatility of ABMs means that they suffer from the ``wilderness of bounded rationality'' \citep{Sims1980} both due to the large number of parameters and the different choices of behavioral rules, and have been described as ``black boxes'' able to generate any phenomena (see \citet{FagioloRoventini2017} for a discussion). 
Consequently, it is imperative to understand the full set of possible generated phenomena how robust they are to changes in parameters. 

Our research provides a first answer to the question of \textit{what are the different types of dynamics that an agent-based model can generate?} 
In the present paper, we first establish which parameter combinations are most influential for the observed dynamics. Interestingly, we find that across models, only few parameter combinations matter, as already alluded to in \citet{GualdiEtAl2015}. Using this feature, we develop an algorithm to systematically uncover the set of possible dynamics that an agent-based model can generate, and apply it to a medium-sized ABM. Our method is able to recover the full set of dynamics in only a few steps.

A model is a mapping from a parameter space to an observable output space. Changing the location in the parameter space changes the observed output. Research by \citet{GutenkunstEtAl2007} has shown that there are generically only a few \textit{stiff} directions in the parameter space that significantly change the observable dynamics\footnote{
	In particular we refer to the seminal work of \citet{GutenkunstEtAl2007}, with follow-on work by \citet{ApgarEtAl2010}, \citet{TranstrumEtAl2010, TranstrumEtAl2011, TranstrumEtAl2015}, \citet{MachtaEtAl2013}, \citet{MannakeeEtAl2016}, and \citet{HsuEtAl2020}
}.
Conversely, there are many \textit{sloppy} directions in which there are no significant changes in the model dynamics. The stiff and sloppy directions are defined as the eigenvectors of the Hessian matrix of a loss function at a point in parameter space. We apply this idea to six macroeconomic models including a medium-sized ABM and five Dynamic Stochastic General Equilibrium (DSGE) models, which represent economists' current ``workhorse'' models for monetary policy \citep{KaplanViolante2018}. We find that each of these models is sloppy: their observable dynamics only change significantly in a handful of directions. These stiff directions typically involve many parameters, and are not aligned with the bare parameter axes, suggesting they may be missed by one-parameter-at-a-time variations. 

In the DSGE models, the key directions are dominated by the parameters governing the exogenous shock processes, while deeper economic parameters typically have little influence. This is a known dependence and limitation of DSGE models \citep{Stiglitz2018}, and reflects the formulation in terms of steady-state deviations for the selected models. In the ABM, the stiffest directions are dominated by parameters relating to \textit{phase transitions} in the model. This finding suggests that common one-parameter-at-a-time sensitivity analysis may not be a sufficiently comprehensive tool to assess model sensitivity.
More sophisticated multi-parameter sensitivity analysis methods exist (see \citet{Ligmann-ZielinskaEtAl2020} for a recent review). However, they have not permeated the macroeconomic ABM literature, and are focused only on specific set policy variables. Our approach complements these methods by considering the sensitivity of the dynamics at a given point in parameter space with respect to all parameters, while remaining computationally efficient. 

We study the implications of the sloppy on the medium-scale Mark-0 ABM of \citet{GualdiEtAl2015, GualdiEtAl2017}, which is a reduced form of the model by \citet{GattiEtAl2011} and has several phase transitions between different dynamics in the model's output.\footnote{A phase transition describes the change in state of a medium, such as the conditions at which water freezes (liquid $\to$ solid). In this case we refer to tipping points, where small changes in parameters drastically alter system dynamics. For example, changing from full employment to full unemployment.} Beginning with the phase where unemployment is constant and high, we find that the stiffest parameter directions represent the direction towards the nearest phase transitions, in this case, the realisation of the model that has a periodic fluctuation around low unemployment. Exploiting this idea, we build an algorithm to step along these stiff directions and thus systematically explore the phase space within a given computational budget. In the case of the Mark-0 model, we show that this approach can efficiently recover all possible dynamics for the unemployment rate for both the low-dimensional and high-dimensional cases. Our method therefore provides a useful tool to navigate the space of parameters to find abrupt changes of behavior and their parameter dependencies.

The issue of searching the parameter space of Agent-based Models has received considerable attention in recent years. For instance, \citep{DosiEtAl2018, DosiEtAl2021} have combined elementary effects and kriging for a global sensitivity analysis.
A prominent method and the closest to ours is meta-modeling (see \citet{LampertiEtAl2018a,vanderHoog2019,ZhangEtAl2020,tenBroekeEtAl2021}, with applications to parameter calibration in \citet{BargigliEtAl2020} and \citet{ChenDesiderio2021}).  
In this approach, observable outputs are sampled at a selection of different points in the parameter space. A simpler model is then constructed that mimics the relationship between the parameters and a given set of outputs, or classifies the outcomes based on some criteria (e.g. using a neural network, as in \citet{LampertiEtAl2018a}).
The benefit of this approach is that once a surrogate model has been constructed, further exploring the parameter space through predicting the class of unseen parameters becomes computationally cheap. 
The crux is that the approach is exploratory only with respect to the classification criteria, such as the existence of a fat tailed distribution of GDP growth, that the modeler chooses to implement.
This requires some ex-ante knowledge of the possible outcomes (or at least the desired outcomes) of the model by which to separate different phases.

Our approach instead is exploratory and agnostic, with no prior assumptions about the dynamics we are seeking, such as to provide an understanding of the full set of possible phenomena that can be generated by the model. We do this through intelligent sampling of the parameter space by exploiting the stiffest directions. An agnostic approach is crucial when some emergent behavior is truly unexpected, as is indeed often the case (see the detailed discussion of this point in \cite{GualdiEtAl2015}). Our method complements the meta-modelling approach, as the discovery of the different phenomena can facilitate a more precise criteria for the classification of parameters by the meta-model. In this way, one could identify new phases using our algorithm, and determine the phase transitions using a meta-model.

To develop the insights mentioned above, we first cover the numerical method for estimating the Hessian matrix from which the parameter directions are extracted (Section \ref{sec:method}). We then apply this to the ABM and five DSGE models introduced in Section \ref{sec:models} to assess their sloppiness (Section \ref{sec:sloppiness}). Finally, we use these insights to develop our algorithm (Section \ref{sec:algorithm}) and test it on the Mark-0 ABM (Section \ref{sec:application}), for which we unveil some behavior that had escaped previous scrutiny.

\section{Extracting Stiff and Sloppy Parameter Combinations}\label{sec:method}
In a first step, we define stiff and sloppy directions in parameter space. 
These directions are given by the eigenvectors of the Hessian Matrix of the change in observable outputs, which represents the relative parameter importance.
Specifically, the eigenvector corresponding to the largest eigenvalue represents the stiffest linear combination of parameters, while the eigenvector corresponding to the smallest eigenvalue is the sloppiest direction.
This approach follows from the work of \citet{GutenkunstEtAl2007}, and was most recently implemented in \citet{HsuEtAl2020} for stochastic models.

More precisely, let $\Phi=\{\Phi_1,\dots,\Phi_P\}$ represent a point in $P$-dimensional parameter space. We quantify the change in observable outcomes due to a shift $\delta$ in parameters from $\Phi$ to $\Phi+\delta$ by the mean-squared loss function:
\begin{equation}\label{eq:square_loss}
    \mathcal{L}(\Phi, \delta) = \frac{1}{2 S K T}\sum_{s}\sum_{k}\sum_{t}\left(\frac{y_{s, k, t}(\Phi+\delta) - y_{s, k, t}(\Phi)}{\norm{y_{s,k}(\Phi)}}\right)^2,
\end{equation}
where $y_{s, k, t}(\Phi)$ is the realisation of output variable $k\in\{1,\dots,K\}$ at time $t\in\{1,\dots,T\}$ for random realisation $s\in\{1,\dots,S\}$. The term $\norm{y_{s,k}(\Phi)}$ is a normalisation to re-scale outputs to the same unit, and is taken to be the mean realisation. To avoid numerical issues, the output variables should all be of a similar order of magnitude. 

The Hessian matrix for the mean-squared loss at point $\Phi$ is defined as
\begin{equation}\label{eq:hessian_def}
    H_{i,j}^{\mathcal{L}}(\Phi) := \left. \frac{d^2\mathcal{L}(\Phi, \delta)}{d\log\Phi_i d\log\Phi_j} \right|_{\delta = 0},
\end{equation}
and we consider parameters $\Phi_i$, $\Phi_j\in\Phi=\{\Phi_1,\dots,\Phi_P\}$, where $P$ is the number of parameters. We take the derivative with respect to the $\log$ parameters to focus on relative parameter changes, as most parameters have different units. Evaluating the Hessian at a predefined target point $\Phi$ yields
\begin{equation}\label{eq:hessian}
    H_{i,j}^{\mathcal{L}}(\Phi) = \frac{1}{S K T}\sum_{s}\sum_{k}\sum_{t}\frac{1}{\norm{y_{s, k, t}(\Phi)}^2} \frac{dy_{s, k, t}(\Phi)}{d\log\Phi_i}\frac{dy_{s, k, t}(\Phi)}{d\log\Phi_j},
\end{equation}
where the second derivative term vanishes as we evaluate $H_{i,j}^{\mathcal{L}}$ at $\delta=0$. This reduces the computational burden to $\mathcal{O}(PS)$ function calls of the underlying model. The target point $\Phi$ is the parameter combination that the modeller wants to investigate. Furthermore, since Eq. \eqref{eq:hessian} is a sum of projectors, it is immediate to see that the numerically determined matrix $H$ is by construction positive definite. 

Due to the simulation based nature of ABMs, we calculate the first-order derivatives numerically using a central difference approach.\footnote{
Central Differences imply $
\frac{dy_{s, k, t}(\Phi^\star)}{d\log\Phi_i} \approx [y_{s, k, t}(\log\Phi^{\star} + \delta e_i) - y_{s, k, t}(\log\Phi^{\star} - \delta e_i)]/{2\delta},
$
where $e_i$ is a vector of zeros such that only element $i$ equals to one, and $\delta$ is the step-size taken in log space. The choice of $\delta$ is arbitrary and potentially model dependent. In our case, we have chosen $\delta=0.1$ for the agent-based models to ensure a sufficiently large step to avoid noise, while remaining small enough to avoid numerical instability. For the DSGE models we found $\delta=0.1$ to be unstable, and so reduced to $\delta=0.001$.
}
While numerical derivatives can be troublesome in sloppy models, we find that the eigenvalues of the Hessian matrix converge to a set of steady ratios as the simulation time $T$ increases, which can be aided by introducing an equilibrating time $T_{\text{eq}}$ before which all observations are dropped. We have generally observed eigenvalues to converge sequentially, with the largest eigenvalue converging first. Consequently, the values of the smallest eigenvalues are likely to be more noisy and closer together.
In contrast to $T$ and $T_{\text{eq}}$, the number of seeds $S$ has little effect on convergence beyond a certain minimum threshold.

To assess the parameter directions of importance, we consider the eigenvectors $\{{v}_1,\dots,{v}_P\}$ of the numerically approximated Hessian matrix (Eq. \eqref{eq:hessian}). The eigenvectors correspond to the sorted eigenvalues $\{\lambda_1,\dots,\lambda_P|\lambda_i\geq\lambda_j~\forall~i<j\}$, where $\lambda_1$ is the largest eigenvalue. The first eigenvector, ${v}_1$, represents the linear combination of parameters corresponding to the stiffest direction in parameter space. 

\section{Macroeconomic Models Considered}\label{sec:models}

\subsection{A simple Agent-Based Model: Mark-0}

The first agent-based model under consideration is the Mark-0 model of \citet{GualdiEtAl2015} that was expanded in \citet{GualdiEtAl2017} and \citet{BouchaudEtAl2018}, with a recent application to the COVID crisis by \citet{SharmaEtAl2020}.\footnote{
We refer to these papers for detailed pseudo-code as well as a mathematical description of the model and its properties. In particular, \citet{SharmaEtAl2020} provide a full description.}
The Mark-0 model is a simplified medium-scale hybrid-ABM of a closed economy with an aggregate household and heterogeneous firms.\footnote{
Hybrid refers here to the characteristic that only one set of agents (firms in the Mark-0 model) are represented by a multiplicity of agents, while other groups (e.g. households) are simply formulated as an aggregate dynamic or representative agent
}
The model can replicate several \textit{macro-states} depending on the parameters  (including, for example, full employment and unemployment, endogenous crises, high-inflation high-output, and low-inflation low-output scenarios).
While Mark-0's aggregate behaviour is most likely not quantitatively precise, it generates qualitatively plausible and generic dynamical behaviours. In the baseline model, \citet{GualdiEtAl2015} have identified four distinct phases for the unemployment rate: full employment (FE), full unemployment (FU), residual unemployment (RU), and endogenous crises (EC).
We chose the Mark-0 model because its' phase diagram has been well-studied in previous work.
This allows us to compare our analysis of key parameter directions and the phase-space exploration algorithm (see Section \ref{sec:algorithm}) to previously studied and explored phases. 
We use the most recent version of \citet{SharmaEtAl2020}, but exclude the central bank, which leaves a fourteen-dimensional parameter space ($P=14$). Table \ref{tab:mark0_parameters} provides an overview of the $P=14$ parameters that we consider in the Mark-0 model, in decreasing order of relevance to this paper. These parameters were predetermined by \citet{GualdiEtAl2015}, and we adopt them as the baseline parameters on which we focus our exploration, unless otherwise stated.

For the purposes of this paper, we focus our attention on only the unemployment rate in the Mark-0 model (equivalent to output in Mark-0) because the phases identified by \citet{GualdiEtAl2015} are most visible in the dynamics of the unemployment rate.\footnote{In Figure \ref{fig:eigenvalue_spectra}(a) we do consider all possible output series when calculating the Hessian. To ensure consistent scales when evaluating all output series, we apply the transformation $\hat{x}_t=\ln(x_t + c)$ with $c=10^7$, to re-scale each of the observable output series $x$. Without this transformation, some variables (e.g. inflation) dominate as the $\norm{y_{s,k}(\Phi^\star)}$ term in Eq. \eqref{eq:square_loss} becomes extremely small.}
For the approximation of the Hessian we use $S=20$ seeds. We set a duration of $T=30,000$ time steps with an initial cutoff of $T_{\text{eq}}=10,000$, after which we find that the first two eigenvalues in the spectrum have converged to a steady state.

\begin{table}[htb!]
\centering
\scalebox{1.0}{
\begin{threeparttable}
\small
\caption{Parameters of the Mark-0 Model without Central Bank sorted in order of their relevance to this paper. The notation follows that of \citet{SharmaEtAl2020}}
\label{tab:mark0_parameters}
\begin{tabular}{lll}
\toprule
\mc{1}{l}{Symbol}&\mc{1}{l}{Default}&\mc{1}{l}{Definition}
\\
\midrule
$\rho^\star$                   & 1\% & Baseline interest rate                       \\
$\Theta$                       & 2.5 & Default threshold                            \\
$\alpha_\Gamma$                & 50 & Loan rate effect on $R$                       \\
$c_0$                          & 0.5 & Baseline propensity to consume               \\
$R$                            & 2 & Hiring/firing rate                             \\
$\frac{\gamma_w}{\gamma_p}$    & 1.0 & Adjustment ratio                             \\
$\gamma_p$                     & 0.1 & Price-adjustment size                        \\
$\eta_0$                       & 0.1 & Baseline firing propensity                   \\
$f$                            & 0.5 & Bankruptcy effect on bank interest rates     \\
$\delta$                       & 2\% & Dividend share                               \\
$\beta$                        & 2.0 & Household intensity of choice                \\
$\tau^R$                       & 0.5 & Weight $\pi^{ema}$ on $\hat{\pi}$            \\
$\phi$                         & 0.1 & Revival frequency                            \\
$\omega$                       & 0.2 & EWMA memory                                  \\
\bottomrule
\end{tabular}
\end{threeparttable}
}
\end{table}

\subsection{Dynamic Stochastic General Equilibrium (DSGE) Models}

To contrast with the Mark-0 model, we also consider several standard DSGE models. Despite multiple shortcomings \citep[see e.g.][]{FagioloRoventini2017, Stiglitz2018}, DSGE models remain the working horse models of many macro-economists \citep{ChristianoEtAl2018, Blanchard2018}. The first three models are borrowed to chapters 2, 3 and 8 of \citet{Gali2015}, and represent the classic real business cycle and new Keynesian frameworks (chapter 8 being an open-economy model). These three models have $P=9$, $12$, and $13$ parameters respectively, and have not been fitted to data explicitly. Rather they form the theoretical backbone of current DSGE models. We also consider three models that have been fitted empirically, namely the canonical \citet{SmetsWouters2007} model with $P=35$ parameters, the financial frictions model of \citet{DelNegroEtAl2015} ($P=50$) and the term-premium model of \citet{CarlstromEtAl2017} ($P=51$). 
For the approximation of the Hessian in the DSGE models we use $S=100$ seeds. We set a duration of $T=5,000$ without an initial cutoff since these models are initialised to their steady state already. The outputs considered are the time-series of the log-deviations from the steady state.

\section{Are Macroeconomic Models Sloppy?}\label{sec:sloppiness}

\begin{figure}[htb!]
    \centering
    \includegraphics[width=0.7\textwidth]{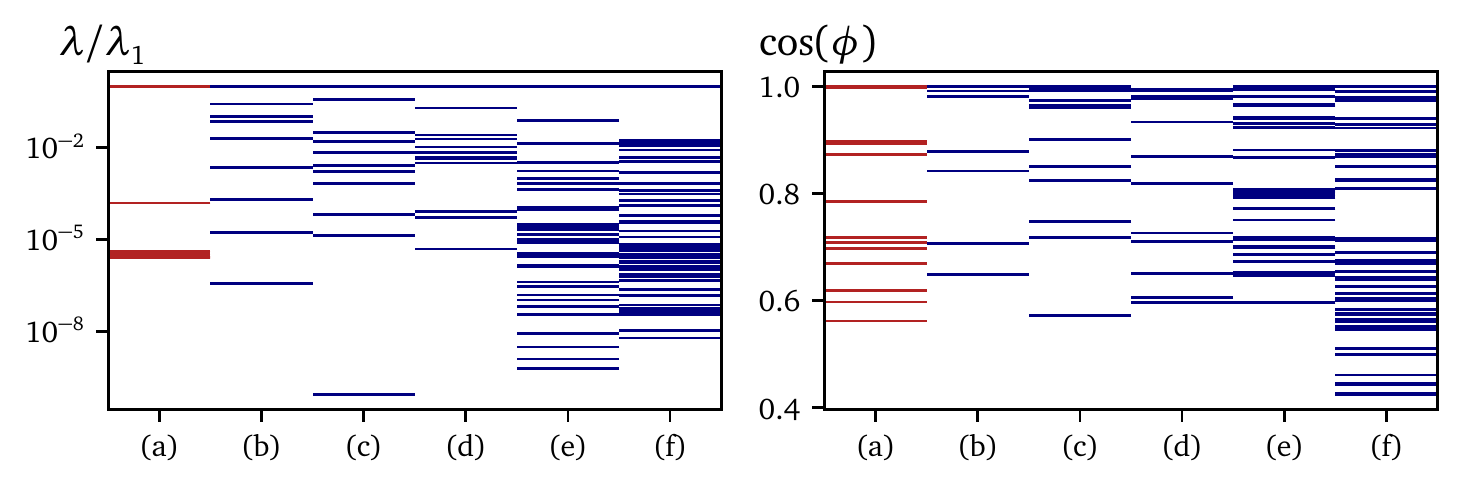}
    \caption{(LHS) Eigenvalue Spectra relative to the largest eigenvalue ($\lambda/\lambda_1$), and (RHS) absolute cosine similarities between each eigenvector and its closest bare parameter axis. Note that the ranking of the cosine similarities does not necessarily correspond to that of the eigenvalues, see Table  \ref{tab:cosine}. The models include in red the Mark-0 ABM \citep{GualdiEtAl2015} with all variables evaluated in the full employment phase (a, $\Theta=2.5$, $\rho^\star=1 \%$), with the remaining parameters from \citet{GualdiEtAl2015}. In blue, the Dynamic Stochastic General Equilibrium models of (b) \citet{Gali2015} chpt. 2, (c) chpt. 3, and (d) chpt. 8, as well as the estimated models of (e) \citet{SmetsWouters2007}, (f) \citet{DelNegroEtAl2015}, and (g) \citet{CarlstromEtAl2017} evaluated at the authors' reported parameters.}
    \label{fig:eigenvalue_spectra}
\end{figure}

Despite their different frameworks and levels of intricacy, all of the models display the sloppy phenomenology found in systems biology models \citep{GutenkunstEtAl2007}. Across all models, we observe that the eigenvalue spectrum spans several decades (Figure \ref{fig:eigenvalue_spectra}). In particular, for the DSGE models (b-g) the distribution is uniform across decades, which is similar to models studied in systems biology. However, the presence of phase transitions in the Mark-0 agent-based model with all observable outputs lead to a different eigenvalue spectrum (a), which is non-uniformly distributed. Instead, there are two deviating eigenvalues and a grouping of the remainder around $10^{-5}$. This occurs for two reasons: First, these stiff eigenvalues correspond to the directions of the closest phase transitions, which cause extreme changes in the model dynamics, while the latter ones cause only minor shifts in the model's steady state or within the phase. The Mark-0 was constructed to be a reduced version of the Mark-1 model \citep{GattiEtAl2011}, such that only parameters affecting the phases remain. The second reason is that the amount of information, $T$, required to further distinguish the small eigenvalues in these models grows with their magnitude (as in \citet{HsuEtAl2020}), such that small eigenvalues may not be accurately estimated and actually be more spread out than shown in Figure \ref{fig:eigenvalue_spectra}.

\begin{table}[hbt!]
\centering
\scalebox{1.0}{
\begin{threeparttable}
\small
\caption{Absolute cosine similarity between eigenvectors and the most aligned bare parameter axes for the first five eigenvectors of (a) the Mark-0 ABM (Gualdi et al., 2015) with all variables evaluated in the full employment phase (a, $\Theta=2.5$, $\rho^\star=1 \%$), and for only the unemployment rate in the full employment phase (FE, and FE II, $\Theta=5.0$, $\rho^\star=1\%$), the residual unemployment phase (RU, $\Theta=2.5$, $\rho^\star=5\%$, and RU II, $\Theta=5.0$, $\rho^\star=5\%$), and the full unemployment phase (FU, $\Theta=2.5$, $\rho^\star=15\%$, and FU II, $\Theta=5.0$, $\rho^\star=12.5\%$). All remaining parameters are from \citet{GualdiEtAl2015}. The Dynamic Stochastic General Equilibrium models of (b) \citet{Gali2015} chpt. 2, (c) chpt. 3, and (d) chpt. 8, as well as the estimated models of (e) \citet{SmetsWouters2007}, (f) \citet{DelNegroEtAl2015}, and (g) \citet{CarlstromEtAl2017} evaluated at the authors' reported parameters.}
\label{tab:cosine}
\begin{tabular}{llllllllllllllll}
\toprule
&\mc{1}{c}{$a$}&\mc{1}{c}{$FE$}&\mc{1}{c}{$FE~II$}&\mc{1}{c}{$RU$}&\mc{1}{c}{$RU~II$}&\mc{1}{c}{$FU$}&\mc{1}{c}{$FU~II$}&\mc{1}{c}{$b$}&\mc{1}{c}{$c$}&\mc{1}{c}{$d$}&\mc{1}{c}{$e$}&\mc{1}{c}{$f$}&\mc{1}{c}{$g$}\\
\midrule
1 & 1.00 & 0.81 & 0.81 & 0.91 & 0.91 & 0.93 & 0.98 & 0.98 & 0.75 & 0.98 & 0.99 & 1.00 & 0.74 \\  
2 & 0.60 & 0.58 & 0.58 & 0.84 & 0.89 & 0.80 & 0.44 & 0.65 & 0.57 & 0.93 & 0.99 & 0.88 & 0.89 \\  
3 & 0.78 & 0.78 & 0.78 & 0.55 & 0.59 & 0.58 & 0.55 & 0.84 & 0.96 & 0.65 & 0.96 & 0.67 & 0.80 \\  
4 & 0.56 & 0.54 & 0.54 & 0.48 & 0.49 & 0.59 & 0.72 & 0.71 & 0.72 & 0.87 & 0.94 & 0.87 & 1.00 \\  
5 & 0.71 & 0.58 & 0.58 & 0.45 & 0.59 & 0.49 & 0.55 & 0.88 & 0.99 & 0.73 & 1.00 & 1.00 & 0.91 \\  
\bottomrule
\end{tabular}
\end{threeparttable}
}
\end{table}

To provide some intuition on the degree of sloppiness, the distance that can be travelled in a direction $v_i$ given by an eigenvector without changing the observed output significantly (as measured by the loss function) is proportional to $1/\sqrt{\lambda_i}$ \citep{GutenkunstEtAl2007}. Thus, for the Mark-0 model (Figure \ref{fig:eigenvalue_spectra}(a)), one must travel roughly $\sim10^{5/2}$ times as far in the sloppiest direction as in the stiffest direction to achieve a change in observable output that is of similar magnitude (in the absence of any phase transitions).
To put this into perspective, if one wants to empirically fit the parameters underlying these models by constraining each direction to within a 10\% confidence interval, one would require $10^5$ times the amount of data for the sloppiest direction in Mark-0 as for the stiffest \citep{MachtaEtAl2013}. In the context of macroeconomics, this implies a high data requirement that is infeasible. However, it may not be necessary for Agent-based Models to be fit to such a high degree of precision in regards to each of their parameters. As noted in \citet{GutenkunstEtAl2007}, ``concrete predictions can be extracted from models long before their parameters are even roughly known'' because observed outputs depend only on a few stiff directions. 


The importance of the stiff directions suggests the question of what the major constituents of these linear combinations are.We first quantify the degree to which the eigenvectors correspond to the bare parameter axes by considering the cosine similarity between the two vectors, where the bare axes represent the natural basis in the P-dimensional parameter space. Specifically, for each eigenvector we report the absolute cosine similarity of the bare axes with which it is most closely aligned ($\cos\phi({v}_i,{v}_j)=\frac{{v}_i\cdot{v}_j}{\norm{{v}_i}\norm{{v}_j}}$).\footnote{The cosine similarity between two vectors is defined by $\cos\phi({v}_i,{v}_j)=\frac{{v}_i\cdot{v}_j}{\norm{{v}_i}\norm{{v}_j}}$. In our case, we compare each eigenvector ${v}_i$ to an identity vector $e_p$ representing parameter direction $p$, where $e_{ip}=0~\forall i\neq p$ and $e_{pp}=1$. From this, we extract the cosine similarity of the eigenvector to the most aligned parameter axis: $sim_i=\max_{p\in P} |\cos\phi({v}_i, e_p)|$.} 
A value of one implies that the given eigenvector contains only a single non-zero parameter entry, rather than a combination of parameters.
Figure \ref{fig:eigenvalue_spectra} (right panel) shows the distribution of the cosine similarities for each eigenvector and its closest bare parameter axis. Note that the ranking here does not correspond to the ranking of eigenvalues; Table \ref{tab:cosine} shows instead the cosine similarities for the stiffest five directions. Many observed directions do not correspond to bare parameter axes, though there is a significant portion that do. We consider these separately for DSGE and ABM models.

\subsection{Stiff Directions in DSGE models}

Based on Figure \ref{fig:eigenvalue_spectra}, DSGE models can be described as sloppy models, with many eigenvalues orders of magnitude smaller than the largest one. This means that many combinations of parameters are in fact quite irrelevant for the dynamics of the system, and that one should be able to construct reduced models with similar performance.

The stiffest parameter directions frequently correspond to individual parameter axes, with cosine similarities close to 1. These directions are often the parameters related to the exogenous shock processes, such as their persistence or co-variances. For the estimated models ($f$, $g$, and $h$ in Figure \ref{fig:eigenvalue_spectra}), almost all of the first 5 stiffest eigenvectors contain large elements relating to the shock processes, with a few also containing a single additional parameter (see Table \ref{tab:cosine}). This reflects the adiabatic nature of these models, which are formulated as deviations from a given steady state, such that shocks in a given period dominate the dynamics of the model. In light of this, once the exogenous processes have been fixed, variations in the further parameters appear to offer comparatively little discernible change in the models' output. 


\subsection{Stiff Parameter Directions and Phase Transitions in Agent-based Models}
\begin{figure}[htb!]
    \centering
    \includegraphics[width=0.8\textwidth]{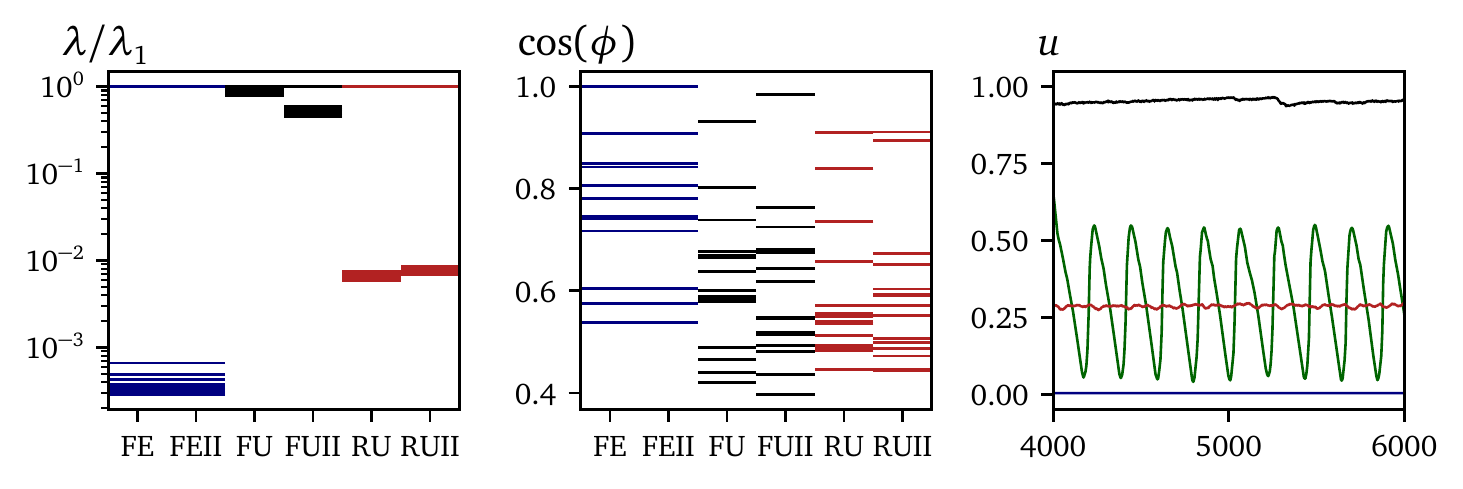}
    \caption{(LHS) Eigenvalue Spectra relative to the largest eigenvalue ($\lambda/\lambda_1$), (Center) the absolute cosine similarities between each eigenvector and its most aligned bare parameter axis, and (Right) the unemployment rate for different phase, for the Mark-0 ABM \citep{GualdiEtAl2015}. At variance with Figure \ref{fig:eigenvalue_spectra} we restrict the estimation to only the unemployment rate as a variable, with parameters in: the full employment phase (blue, with FE, $\Theta=2.5$, $\rho^\star=1\%$, and FE II, $\Theta=5.0$, $\rho^\star=1\%$), the residual unemployment phase (red, with RU, $\Theta=2.5$, $\rho^\star=5\%$, and RU II, $\Theta=5.0$, $\rho^\star=5\%$), and the full unemployment phase (black, FU, $\Theta=2.5$, $\rho^\star=15\%$, and FU II, $\Theta=5.0$, $\rho^\star=12.5\%$). The EC phase (green) is evaluated at $\Theta=1.3$, $\rho^\star=1\%$. All remaining parameters are from \citet{GualdiEtAl2015}.}
    \label{fig:eigenvalues_mark0}
\end{figure}
In the Mark-0 agent-based model, the stiffest directions are dominated by parameters relating to phase transitions in the model. To explore this, we focus on the phases of the unemployment rate: full employment (FE), residual unemployment (RU), endogeneous crises (EC), and full unemployment (FU) (see Figure \ref{fig:eigenvalues_mark0}). The eigenspectra for two points in three phases are shown in Figure \ref{fig:eigenvalues_mark0}. They can generally be described by an outlier first eigenvalue, followed by a dense body of the remaining eigenvalues. The degree of separation between the first and second eigenvalue depends on the phase considered, with a large deviation in the FE phase, and a small one in the FU phase. 

\begin{figure}[htb!]
    \centering
    \includegraphics[width=0.7\textwidth]{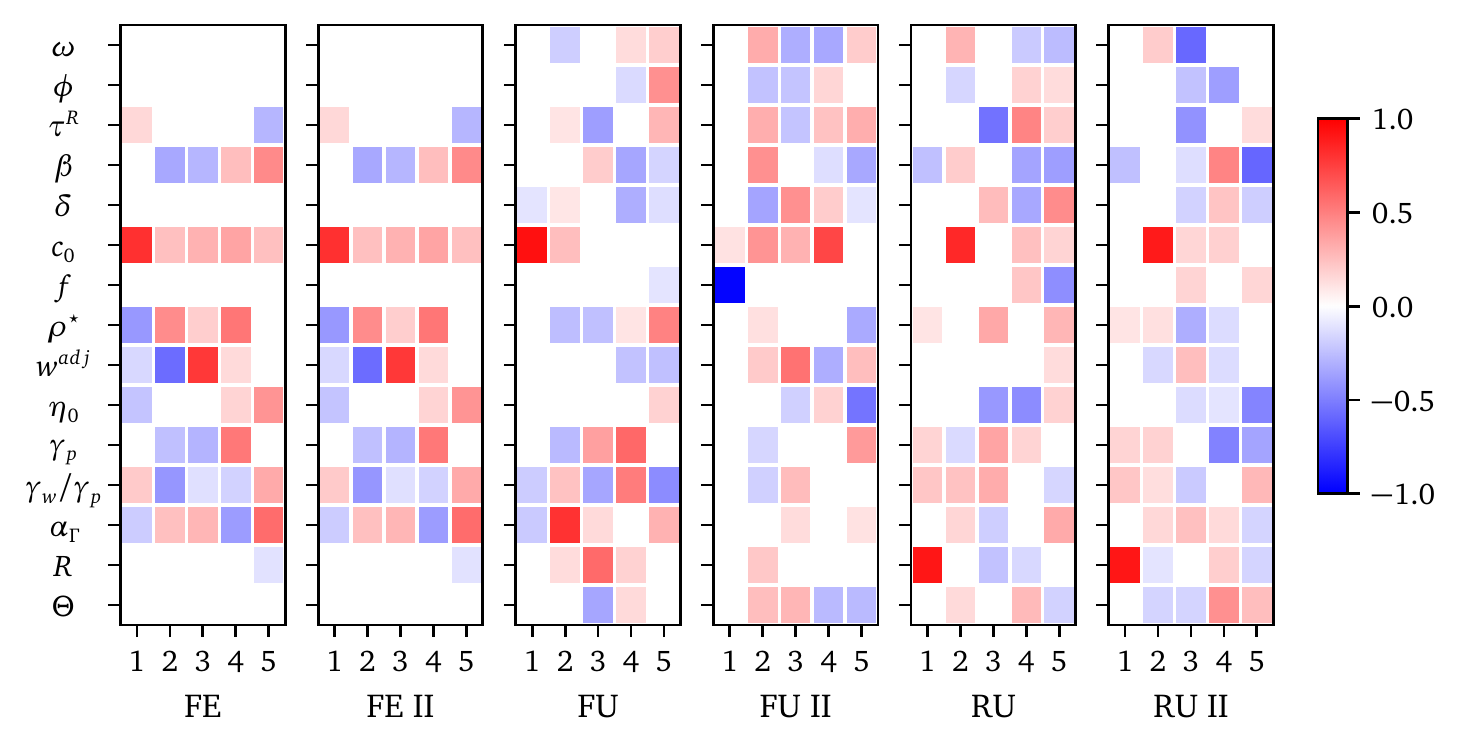}
    \caption{Heatmaps of the components of the stiffest 5 eigenvectors of the Mark-0 ABM focused on the unemployment rate with all parameters. The sign of the eigenvector is such that the angle with the direction $c_0$ is positive. All components with a magnitude between -0.1 and 0.1 have been set to zero for exposition purposes. The directions were evaluated in the full employment phase (FE, $\Theta=2.5$, $\rho^\star=1\%$, and FE II, $\Theta=5.0$, $\rho^\star=1\%$), the residual unemployment phase (RU, $\Theta=2.5$, $\rho^\star=5\%$, and RU II, $\Theta=5.0$, $\rho^\star=5\%$), and the full unemployment phase (FU, $\Theta=2.5$, $\rho^\star=15\%$, and FU II, $\Theta=5.0$, $\rho^\star=12.5\%$). All remaining parameters are from \citet{GualdiEtAl2015}.}
    \label{fig:eigenvectors}
\end{figure}

\begin{figure}[htb!]
    \centering
    \includegraphics[width=0.9\textwidth]{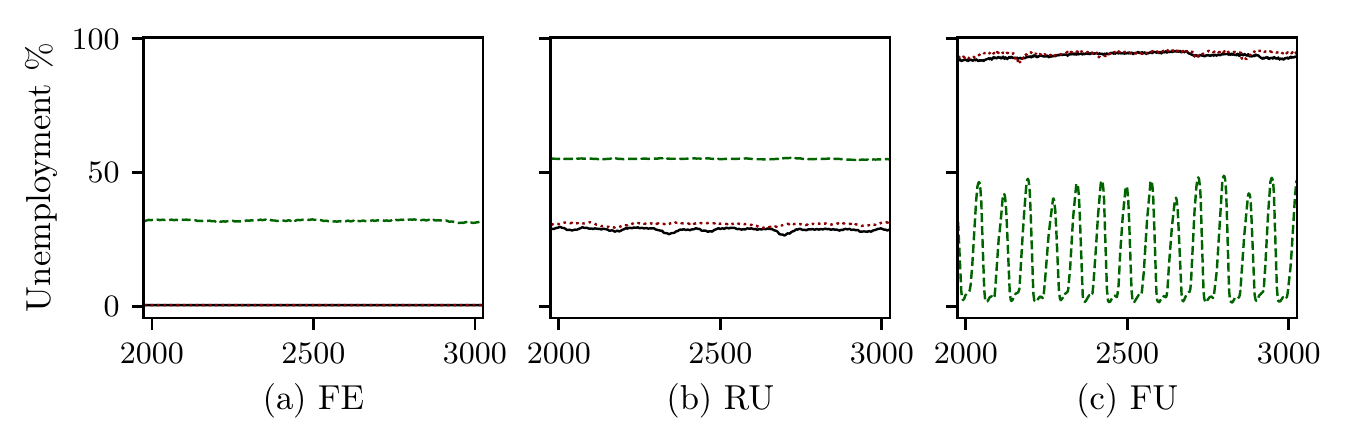}
    \caption{Unemployment dynamics in different phases in their original form (black), with a unit log step in the stiffest (green) and sloppiest (red) directions identified in Figure \ref{fig:eigenvectors}, i.e. $\log\Phi'=\log\Phi + v_i$. The panels are: (left) the full employment phase (FE, $\Theta=2.5$, $\rho^\star=1\%$), (center) the residual unemployment phase (RU, $\Theta=2.5$, $\rho^\star=5\%$), and (right) the full unemployment phase (FU, $\Theta=2.5$, $\rho^\star=15\%$). All remaining parameters are from \citet{GualdiEtAl2015}.}
    
    \label{fig:phases}
\end{figure}

To understand the components of the stiffest directions, Figure \ref{fig:eigenvectors} shows a heatmap of the stiffest five eigenvectors for the three phases considered. In each case, the direction for the first eigenvector is dominated by one of the variables which have been found to lead to phase transitions in the ``handcrafted`'' analysis of \citet{GualdiEtAl2015}, including the parameters $R$ and $\alpha_\Gamma$ (for a definition of these variables, see Table \ref{tab:mark0_parameters}). In addition, we have here also identified $c_0$ to be a key variable in the phase transition, missed in the analysis of \citet{GualdiEtAl2015}. While the stiff directions are dominated by a single value, they are nonetheless unaligned with the bare parameter axes, as shown by the non-unity cosine similarities. In fact, the combination of parameters is generally one of multiple other phase-relevant parameters. For example, in the RU phase these mixture parameters include $\beta$, and $\gamma_w$ and $\rho^\star$, each of which are related to varying phase transitions.
To illustrate the relation between stiff directions and phase transitions, Figure \ref{fig:phases} shows the dynamics of the unemployment rate in the Mark-0 model starting in each phase (black lines), with $\log$ step in the stiffest (green) and sloppiest (red) directions (i.e. $\log\Phi'=\log\Phi + v_i$). Note that for the full employment case (left panel), we observe a transition to residual unemployment, while for the full unemployment phase (right panel) we observe a transition to the endogenous crises phase. For the residual unemployment phase there is no explicit transition, but a strong move towards full unemployment. Note that in all cases, there is almost no change in the dynamics when traversing the sloppy direction with an equal step size. The key observation is that the variables identified in the stiffest directions do indeed point towards nearby phase transitions.\footnote{Note however that eigenvectors are unsigned, so following such directions can lead towards the phase transition or away from it, but at the fastest possible pace.} These results suggest that looking to the stiffest direction is a means to identify their different phases and the boundaries separating them.

\subsection{Transitions in Two Dimensions}

\begin{figure}[htb!]
    \centering
    \includegraphics[width=1.0\textwidth]{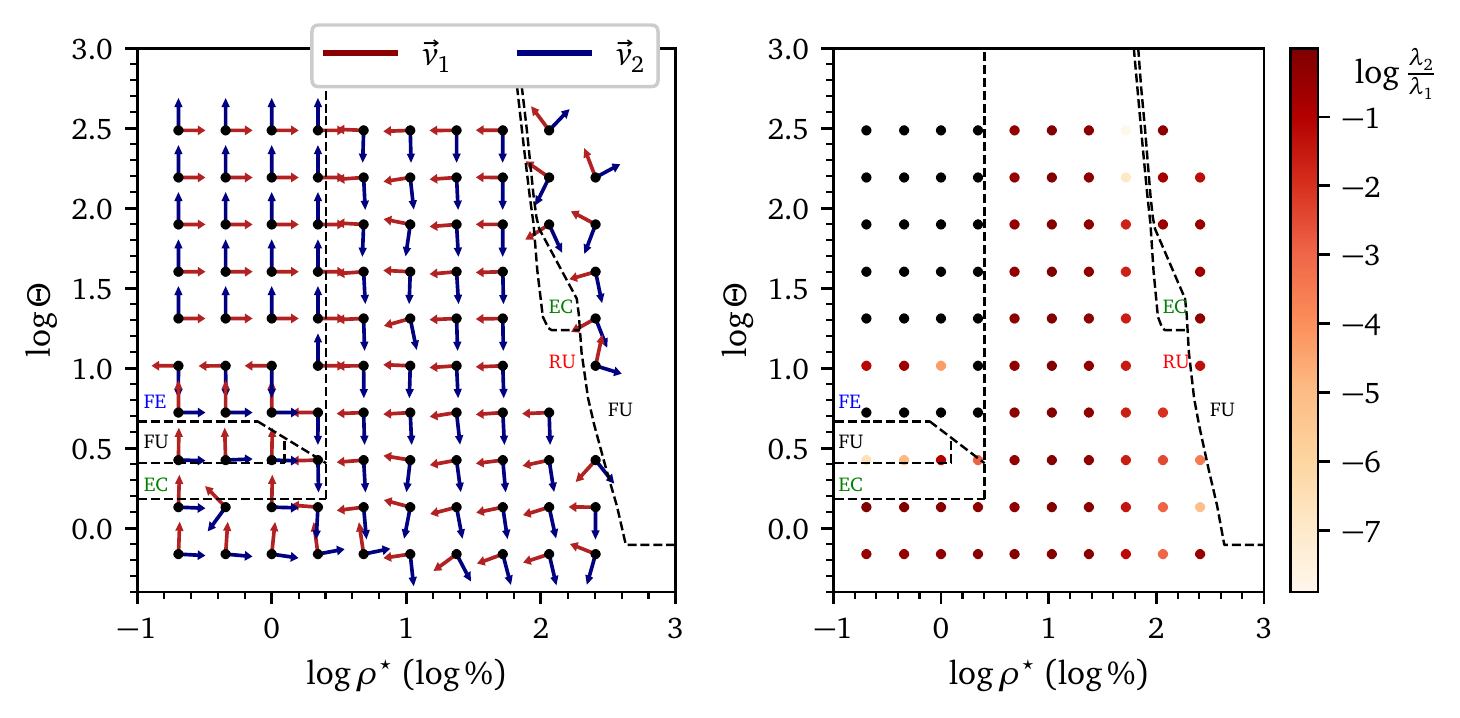}  
    \caption{(LHS) Phase diagram of the Mark-0 model in the $(\log\Theta,\log\rho^\star)$ plane with projected eigenvectors. The eigenvectors, ${v}_1$ (red) and ${v}_2$ (blue), are based on the 2-dimensional Hessian with only $\Theta$ and $\rho^\star$ using unemployment as output variable. Dashed black lines indicate the locations of phase transitions. (RHS) logarithm of the ratio of the first to the second eigenvalue at different points in the phase space. The black dots indicate $\ln(\lambda_2/\lambda_1)<-10$. Simulation parameters: $T=30,000$, $S=20$, $T_{\text{eq}}=10,000$. The parameters of the model, other than $\Theta$ and $\rho^\star$, are from \citet{GualdiEtAl2015}}
    \label{fig:mark0_phases}
\end{figure}

For the unemployment rate dynamics in the two dimensional plane (bankruptcy threshold $\Theta$, central bank's baseline interest rate $\rho^\star$), we find that the stiffest direction consistently points in the direction of a nearby phase transition. Figure \ref{fig:mark0_phases} shows the approximate phase diagram in the $(\log\Theta, \log\rho^\star)$ space (dashed black lines), together with the estimation of the first (red) and second (blue) eigenvectors for a grid of points.

The phase diagram (Figure \ref{fig:mark0_phases} left panel) suggests that the line formed by the first eigenvector is also generally the shortest path to the phase transition. The second eigenvector then points in the direction of the second closest phase transition or parameter boundary. This is clearest when considering the intersection of FE, RU and FU near $\log\Theta\approx0.5$ and $\log\rho^\star\approx0$. Here, at $(0,0.4)$ the stiffest direction is the closer FU$\to$FE transition, while for $(0.2, 0.6)$ the stiffest direction is the closer FE$\to$RU transition. While this is true near phase boundaries, it is also true further away from them. In the FE phase, far from the boundary, the stiffest direction still points towards the closer FE$\to$FU transition. 

The proximity of a phase transition appears, as one would expect, to lead to a strong splitting in the eigenvalues. The right panel of Figure \ref{fig:mark0_phases} shows the ratio of the first ($\lambda_1$) and second ($\lambda_2$) eigenvalues corresponding to the grid in the left panel. Near the FE$\to$RU transition as well as the RU$\to$FU transitions the ratio $\lambda_2/\lambda_1$ tends to zero. This is likely due to the high value $\mathcal{L}$ generated by a transition over the phase boundary, which eclipses other changes. On the other hand, it is also possible that this is simply a result of a drop in importance of all other parameters, as in the FE phase the second direction is irrelevant for almost all parameter values (except near phase transitions). 

The implication of these findings is twofold: first, the stiff eigenvectors are indicators for the shortest path to the closest phase transition, second, the relationship between the eigenvalues is indicative of the distance to the phase transition, and potentially the number of different transitions.

\section{A New Method for Systematic Phase Exploration}\label{sec:algorithm}

Using the formalism above, we now construct a probabilistic algorithm that exploits the insight that the stiffest parameter directions indicate the direction of the largest change in observed dynamics in order to generate a maximally diverse sampling of the observable output space at a relatively low computational cost. The general principle of the algorithm is to evaluate the stiffest direction at a given point, travel along them to a new point, and repeat the exercise for a given number of steps. We are constructing a new intelligent means to sample the parameter space, compared to existing methods such as Sobol sampling or Nearly Orthogonal Latin Hypercubes.

\subsection{Formulation of the Algorithm}
To begin with, the modeller must decide a number of steps $N$ for which the algorithm should run, as well as a starting point $\Phi(0)$ in parameter space. Furthermore, the modeller should set the hyper-parameters for the Hessian estimation: the number of random realisations $S$, the simulation time $T$ and the equilibrating time $T_{\text{eq}}$, such that the first two eigenvalues converge to a steady ratio. The algorithm then proceeds sequentially for each step $n\in \{1,2,\dots,N\}$ as follows: 

\begin{enumerate}
    \item Evaluate the Hessian matrix $H^{\mathcal{L}}(\Phi(n))$ and decompose it into its eigenvalues and associated eigenvectors. For the purpose of the algorithm, we consider only the first two eigenvectors, ${v}_1$ and ${v}_2$, together with their respective eigenvalues $\lambda_1$ and $\lambda_2$. More can be considered, but an accurate estimation of their spectrum requires an exponentially increasing simulation time $T$ (see e.g. \citet{HsuEtAl2020}).
    \item Select the direction, ${v}(n)$ of the next step via:
    \begin{equation}
        {v}(n) = 
        \begin{cases}
            {v}_1, & \text{with probability } \lambda_1 / (\lambda_1+\lambda_2) \\
            {v}_2, & \text{with probability } \lambda_2 / (\lambda_1+\lambda_2)
        \end{cases},
    \end{equation}
    such that $\lambda(n)$ is the corresponding eigenvalue. This selection generally steps in the stiffest direction, unless both directions are important, as could be the case in the presence of multiple close phase transitions for instance. We find that the first two eigenvectors are sufficient to span the search space and to find all important dynamics of the Mark-0 model (can also be seen in Figure \ref{fig:eigenvalue_spectra}).
    \item Since the sign of the eigenvector returned by the algorithm is ill-defined, ensure that the direction is consistent, that is
    \begin{equation}
        {v}(n) = 
        \begin{cases}
            {v}(n), & \text{if } \arccos({v}(n)\cdot{v}(n-1)) \leq \ang{165} \\
            -{v}(n), & \text{otherwise}
        \end{cases},
    \end{equation}
    which prevents the algorithm from oscillating back and forth between two points across the same phase transition.\footnote{It is possible to choose the sign of the eigenvector randomly. This non-ballistic motion might lead to a denser exploration of the phase space that is less sensitive to the initial conditions and choice of direction.}
    
    In case this is the first step (i.e. ${v}(n-1)$ does not exist) we determine the direction of the first step by taking a step in both the ${v}(n)$ and $-{v}(n)$ directions, yielding parameter candidates $\Phi(n)^{+}$ and $\Phi(n)^{-}$ (see step 4 for the distance) and picking the candidate with the largest loss relative to the initial parameter point $\Phi(0)$ (see Eq. \eqref{eq:square_loss}).
    
    \item Pick the next point in parameter space by travelling a distance $d$ in the ${v}(n)$ direction which is neither too large nor too small: 
    \begin{equation}\label{eq:steps}
        d = \min\left(\frac{1}{\sqrt{\lambda(n)}}\max\left({\varepsilon}, \varepsilon_\min\sqrt{{\lambda_1}}\right), \varepsilon_\max\right)
    \end{equation}
    \begin{equation}
        \log\Phi^{n+1} = \log\Phi(n) + d\cdot{v}(n)
    \end{equation}
    where the distance depends on three given parameters: $\varepsilon_\min=0.3$, the minimum step-size to take in log-parameter space, $\varepsilon=0.1$, the distance to travel relative to the span of the interval over which the direction can vary ($\sqrt{\lambda(n)}$), and $\varepsilon_\max=1.0$, the maximum distance that can be traversed (corresponds to a factor of at most $e$ in bare parameter space). We introduce $\varepsilon_{\min}$ and $\varepsilon_\max$ to counteract extreme eigenvalues, which may lead to extreme steps (unrealistically large parameters) or extremely small steps such that a large number of steps is required to generate sufficient changes. The goal is to maximise the local exploration while remaining in general proximity of the initial parameters, where our approximation in Eq. \ref{eq:hessian} still holds. For smooth models, the eigenvectors and eigenvalues span a hyperribbon that encloses a region with the same value of the loss function, see \cite{GutenkunstEtAl2007}. Therefore, we make a step with our algorithm to the edge of the hyperribbon.
\end{enumerate}

The most likely path followed by this algorithm is to select the top eigenvector at each step, in this way, the algorithm continually explores the direction of the largest change. Only when the first eigenvalues are similar is there a chance of bifurcating. This implies that in close proximity to a phase transition, where the eigenvalues typically diverge from one another, the direction of the phase transition is selected. Meanwhile, in areas of the parameter space where there are no immediate transitions and the eigenvalues take on a more uniform distribution, it is more likely that the algorithm explores parameter combinations along the second eigenvector as well. It is not critical that the model shows a first order phase transition, a continuous transition to another phase can also detected with the algorithm.

\subsection{Computational Cost}
The computational expense of agent-based models implies that there is typically an upper limit on the number of function calls that can be made. In regards to the algorithm presented, the most costly step is the calculation of the Hessian matrix, which requires $\mathcal{O}(SP)$ function calls. Consequently, the algorithm overall requires $\mathcal{O}(NSP)$ simulations of the agent-based model. For the two-dimensional Mark-0, this implies at least $8\times20\times2\times2=640$ evaluations with a duration of $T=30,000$ time-steps are required (note one factor of two emerges from taking central difference derivatives). For reference, a Hessian matrix for the Mark-0 with $P=14$ parameters, running $S=20$ seeds in parallel, and $T=30,000$, requires $\sim45$ minutes of computation time on an AMD Ryzen 9 5950X processor with 32GB RAM, while a $P=2$ Hessian requires $\sim6$ minutes. A natural benchmark for our computation is the intelligent Latin Hypercube Sampling strategy \citep{McKayEtAl1979, CioppaLucas2007}.\footnote{For Latin Hypercubes, the parameter space is divided into equally probable intervals and then sample points are placed such that there is only one sample in each axis-aligned hyperplane} For the two-dimensional Mark-0, we find that $\sim29$ samples of the parameter space are required to obtain all four phases. Considering $S=20$ random samples to mitigate any noise effects, this corresponds to $29\times20=580$ function calls, which is on par with a seven-step estimation of the algorithm. As the dimension of the parameter-space increases, we expect the number of required parameter samples for the hypercube sampling to increase at a higher rate than the algorithm presented here. Simultaneously, the information gained from the Hessian matrices may prove more valuable than a quasi-random sampling.

\section{An Application to the Mark-0 Agent-Based Model}\label{sec:application}
The testing bed for the algorithm is the identification of the phases of unemployment dynamics in the Mark-0 model, for which the number of parameters is $P=14$. We find that in two, three and sixteen dimensions, our algorithm can recover all the major phases of unemployment (full employment, full unemployment and residual unemployment) within eight steps only. Due to the small parameter region for which the EC phase exists, it is often the case that it is missed in the two-dimensional case. Thus, the recovery of the endogenous crisis phase is dependent on the starting point.
Despite dependence on initial conditions, in almost all cases our algorithm recovers at least one phase that is different to the phase of the initial parameter choice.

\begin{figure}[htb!]
    \centering
    \includegraphics[width=0.9\linewidth]{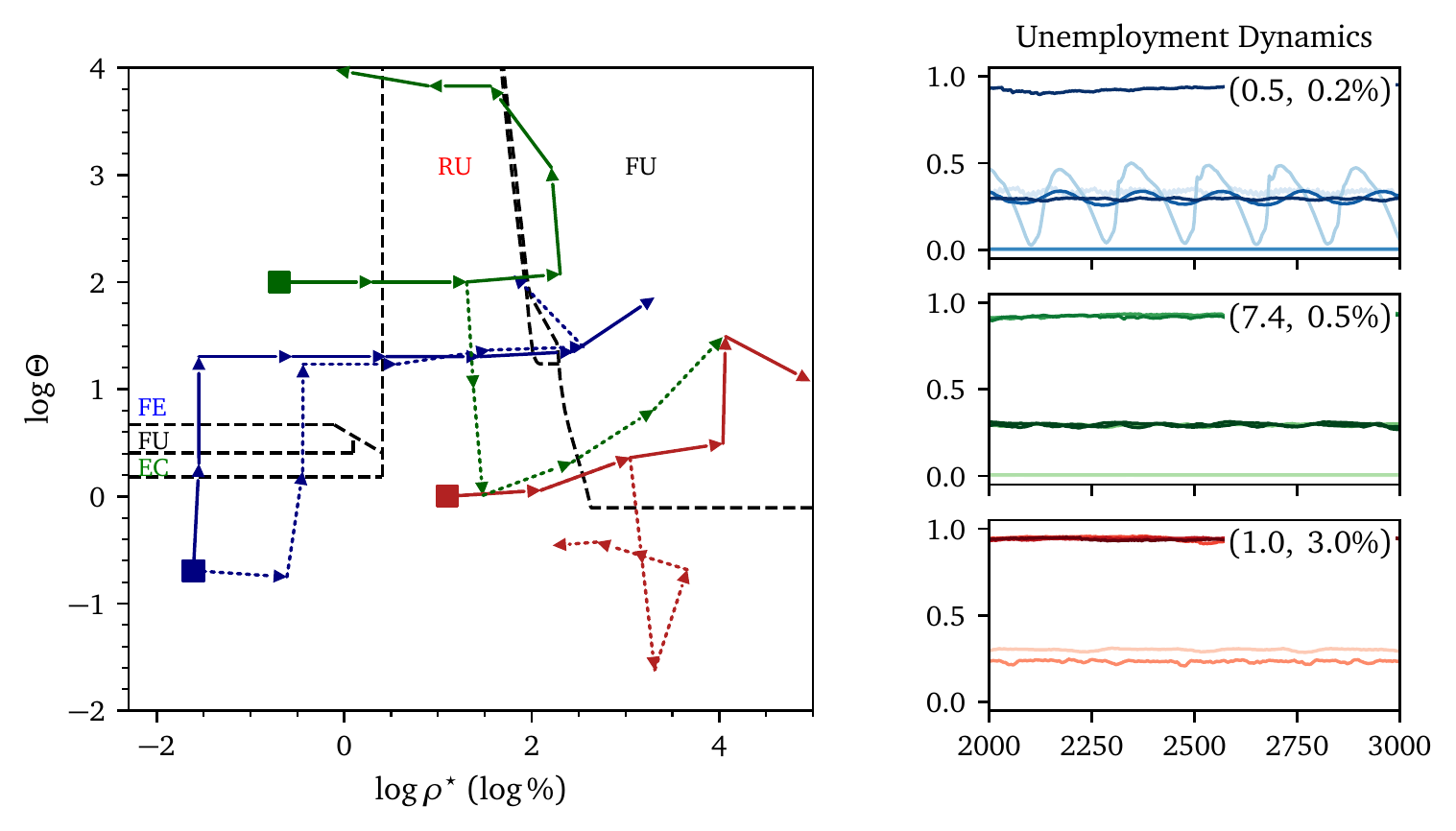}
    \caption{(LHS) Phase diagram of the Mark-0 model without central bank in the $(\log\Theta,\log\rho^\star)$ plane with two likely algorithm paths for three different starting values. Solid lines show the most likely path following only ${v}_1$, while dotted lines indicate an alternate path with mixed steps. Dashed black lines indicate the locations of phase transitions. (b) Dynamics of the unemployment rate for different steps along the ${v}_1$ path for each starting point. Simulation parameters: $T=30,000$, $S=20$, $T_{\text{eq}}=10,000$. The parameters of the model, other than $\Theta$ and $\rho^\star$, are those of \citet{GualdiEtAl2015}. Algorithm parameters are $\varepsilon_\min=0.3$, $\varepsilon=0.1$, $\varepsilon_\max=1.0$.}
    \label{fig:mark0_2d_walks}
\end{figure}

In the two dimensional $(\Theta,~\rho^\star)$-space we find that the algorithm covers at least two phases irrespective of the starting point or the choice of path taken. Figure \ref{fig:mark0_2d_walks} shows two walks originating from each of three different starting points in the $(\Theta,~\rho^\star)$ phase diagram. The solid lines represent the most likely walk in which the algorithm exclusively follows the stiffest direction, ${v}_1$, at each point. The dashed lines represent an alternative random walk, which contains at least one deviation in the direction of the second stiffest direction. The sequence of choices, and the probabilities of choosing the stiffest direction at each point are shown in Table \ref{tab:probabilities}. 

\begin{table}[htb!]
\centering
\scalebox{1.0}{
\begin{threeparttable}
\small
\caption{Probability of choosing eigenvector $v_1$ at each step in three paths taken from a starting point in the $(\Theta,\rho^\star)$-space. Simulation parameters: $T=30,000$, $S=20$, $T_{\text{eq}}=10,000$. The parameters of the model, other than $\Theta$ and $\rho^\star$, are those of \citet{GualdiEtAl2015}. Algorithm parameters are $\varepsilon_\min=0.3$, $\varepsilon=0.1$, $\varepsilon_\max=1.0$.}
\label{tab:probabilities}
\begin{tabular}{llrrrrrrr}
\toprule
\mc{2}{c}{}&\mc{7}{c}{$\mathbb{P}(\vec{v}^n=\vec{v}_1)$}\\
\cmidrule{3-9}
\mc{1}{c}{Start $(\Theta, \rho^\star)$}
&\mc{1}{c}{Path}&\mc{1}{c}{1}&\mc{1}{c}{2}&\mc{1}{c}{3}&\mc{1}{c}{4}&\mc{1}{c}{5}&\mc{1}{c}{6}&\mc{1}{c}{7}\\
\midrule
$(0.5,~0.002)$ & 1111111 & 0.53 & 0.70 & 1.00 & 1.00 & 0.98 & 0.62 & 0.64 \\  
$(0.5,~0.002)$ & 2111111 & 0.53 & 0.52 & 0.86 & 1.00 & 0.65 & 0.68 & 0.57 \\  
$(1.0,~0.030)$ & 111111   & 0.51 & 0.91 & 0.68 & 0.94 & 0.60 & 0.91 & - \\  
$(1.0,~0.030)$ & 1121111  & 0.51 & 0.91 & 0.68 & 0.60 & 0.63 & 0.57 & 0.94 \\  
$(7.4,~0.005)$ & 1111111 & 1.00 & 1.00 & 0.56 & 0.69 & 0.66 & 1.00 & 0.52 \\  
$(7.4,~0.005)$ & 1122112 & 1.00 & 1.00 & 0.56 & 0.58 & 0.62 & 0.87 & 0.75 \\  
\bottomrule
\end{tabular}
\end{threeparttable}
}
\end{table}

The effectiveness of the algorithm is dependent on its initial position within the parameter space.\footnote{This could potentially mitigated by randomising the first direction, or simply walking in both directions} The most effective walk begins near the parameter boundaries at $\Theta=0.5$ and $\rho^\star=0.2\%$, and covers each of the unemployment phases (see the dynamics in the top right panel of Figure \ref{fig:mark0_2d_walks}). The walk beginning in FE (green) covers the three major phases of FE, RU, and FU within three steps, while missing out on the EC phase. This is likely due to the fact that within this two-dimensional space, the EC phase is only present in small pockets, such that it is easily missed. Finally, the second RU walk (red) is the least effective, as it recovers only the RU and FU phases before terminating at extreme $\rho^\star$ values. 

The probability of following the stiffest direction (Table \ref{tab:probabilities}) reflects the observations made in Section \ref{sec:sloppiness}, that the eigenvalues are drawn apart in the presence of a discontinuous phase transition. In particular, the probability of taking the stiffest direction tends to one when there is a singular close phase transition, while it tends to 50\% when there is either two proximate phase transitions or no proximate phase transitions. We note that the FE phase may be an exception to this, as the second eigenvalue becomes vanishingly small throughout the phase. This may be due to the low-fluctuation steady-state nature of this phase. 

While the full recovery of phases may depend on starting values, each walk identified at least two phases. This suggests that a potentially optimal exploration may require a combination of parameter space sampling combined with short iterations of the algorithm. Alternatively, since the direction of the eigenvectors ${v}_i$ is indeterminate, an alternative option (when not close to parameter boundaries) would be to start the algorithm in both possible directions. In that case, starting points in the central RU phase would explore both the FE and FU phases. 

\subsection{A Tentative Extension to Higher Dimensions}

\begin{figure}[htb!]
    \centering
    \includegraphics[width=0.8\textwidth]{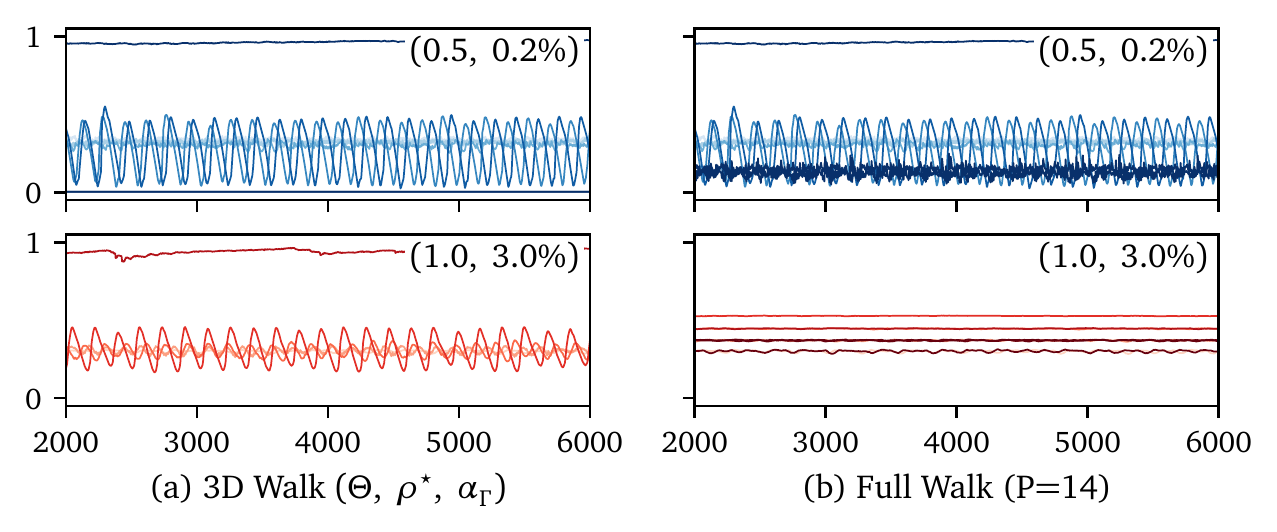}
    \caption{Dynamics of the unemployment rate for different steps in the 3D case ($\Theta,~\rho^\star,~\alpha_\Gamma$) and the multi-dimensional $(P=14)$ case. We consider again the three most likely paths with 5 steps. The initial parameters of the algorithm in the $(\Theta, \rho^\star)$-space are shown in the top right-hand corner of each panel and correspond to those of Figure \ref{fig:mark0_2d_walks}). Simulation parameters: $T=30,000$, $S=20$, $T_{\text{eq}}=10,000$. All other parameters are those of \citet{GualdiEtAl2015}.}
    \label{fig:mark0_high_dim}
\end{figure}

The principal aim of our algorithm is to facilitate phase exploration of agent-based models in high-dimensional parameter spaces. We apply the algorithm to both the three dimensional and fourteen dimensional (all parameters) formulation of the Mark-0 model to identify the unemployment phases. Figure \ref{fig:mark0_high_dim} shows the unemployment rate dynamics for two walks following exclusively the stiffest direction in the $P=3$ (left) and $P=14$ (right) cases. The starting points correspond to those of the two dimensional exploration (Figure \ref{fig:mark0_2d_walks}).

For the previously successful walk beginning at $\Theta=0.5$, $\rho^\star=0.2\%$ we again find that in the $P=3$ case all phases have been recovered, while the full employment phase was missed in the $P=14$ case. For the $\Theta=1.0$, $\rho^\star=3.0\%$ case we find that in the $P=3$ case we obtain the same results as in the $P=2$ case, while in the $P=14$ case there is only a series of level shifts within the residual unemployment phase.\footnote{Note that increasing the step size $\varepsilon$ to 0.3 leads to a wider array of levels explored between the RU and FU phases, at the cost of missing the EC and FE phases in the $\Theta=0.5$, $\rho^\star=0.2\%$ case} In particular, the lower right panel suggests that no different phases are identified (such as the large level shifts between RU and FE, or different oscillatory behaviours between EC and RU). The reason behind this is the agnostic mean-squared loss function (Eq. \eqref{eq:square_loss}) which directly compares the unemployment time-series. This means that qualitatively small changes such as level shifts may dominate. One example is the small shift in periodicity of the EC phase (Figure \ref{fig:mark0_high_dim} top panels) that would lead to a large loss as $T$ becomes large, though it is qualitatively still an EC phase. This suggests that our algorithm should be improved along several dimensions: i) considering an alternative agnostic loss function such as the Kullback-Leibler divergence, ii) including more observables in the loss function, allowing one to better discriminate between different phases; iii) adding an element of randomness in the amplitude and direction of the random walk would remove the (deterministic) dependence on the initial condition. We leave these extensions for further investigations. 

The introduction of further dimensions leads to a different picture on key parameters, as well as the effective dimensions of the model. The introduction of $\alpha_\Gamma$ ($P=3$) leads to a strong representation of $\alpha_\Gamma$ in the stiffest direction, in place of $\rho^\star$. This is because $\alpha_\Gamma$ controls the vertical boundary between the FE and RU phases shown in Figure \ref{fig:mark0_2d_walks}, such that the distance to the phase transition in the $\log\alpha_\Gamma$ direction is likely shorter than in the $\log\rho^\star$ direction. A similar situation occurs also in the $\Theta=1.0$, $\rho^\star=3.0\%$ case, where the introduction of $\alpha_\Gamma$ leads to less frequent changes in $\rho^\star$. This shift is also reflected in the $P=14$ case. The parameters with the largest changes ($R$, $\alpha_\Gamma$, $\beta$, $\Theta$, $\gamma_w/\gamma_p$) are those that have previously been identified to induce changes in phases \citep{GualdiEtAl2015, GualdiEtAl2017, BouchaudEtAl2018}. Meanwhile, for both starting points (Figure \ref{fig:mark0_high_dim}) at least five parameters were left unperturbed, suggesting that the model phases vary in a $Q=9$-dimensional space, suggesting one potential avenue for the reduction of the model. 

\section{Conclusion}
In our paper, we have shown that both agent-based models and DSGE models have a sloppy phenomenology, where their dynamics depend only on a few key stiff parameter directions, much like many models in all fields of science \citep{QuinnEtAl2021}. In the case of the Mark-0 agent-based model, the stiff parameter directions point towards close phase transitions. Exploiting these key directions, we developed an algorithm to systematically explore the phase space of agent-based models, which should apply universally (perhaps at the cost of tweaking the exploration parameters defined in Eq. \eqref{eq:steps}). Applying this algorithm to the Mark-0 model, we have recovered the four phases that were identified in the prior literature, both in low and high-dimensional cases. This method of exploration also appears to be more computationally efficient as the number of parameter dimensions grows.

Since agent-based models are able to generate a rich phenomenology, this tool may aid in understanding the possible scenarios in an agent-based model, as well as their robustness to changes in the underlying parameterisation. This suggests several avenues of further exploration, including extending our analysis to a wider variety of observable outcomes and more complex agent-based models, as well as potentially combining it with a meta-modeling approach to not only infer the set of phases but also their boundaries. Another direction, suggested to us by J. Sethna, would be to follow sloppy directions instead of stiff directions, with the aim of removing irrelevant (combination of) parameters and constructing minimal models in a systematic way once the different behaviors have been classified, along the lines of \citet{QuinnEtAl2021}. Building ``reduced form'' models using such a systematic procedure appears to us as a particularly exciting perspective. 

\section{Acknowledgements}
We thank the members of the EconophysiX lab, as well as Francesco Zamponi for their comments and suggestions. We also James P. Sethna for his thoughtful comments and suggestions. Finally, we thank Davide Luzzati and the Editors for their time and effort. This research was conducted within the Econophysics \& Complex Systems Research Chair, the latter under the aegis of the Fondation du Risque, the Fondation de l’Ecole polytechnique, the Ecole polytechnique and Capital Fund Management. Karl Naumann-Woleske also acknowledges the support from the New Approaches to Economic Challenges Unit at the Organization for Economic Cooperation and Development (OECD).

\newpage
\bibliographystyle{agsm}
\bibliography{RefsSloppyModels.bib}
\theendnotes
\end{document}